\documentclass[12pt,preprint]{aastex}

\newcommand{\degrees}{\mbox{$^\mathrm{o}$~}}
\def\gsim{\;\rlap{\lower 2.5pt\hbox{$\sim$}}\raise 1.5pt\hbox{$>$}\;}
\def\lsim{\;\rlap{\lower 2.5pt\hbox{$\sim$}}\raise 1.5pt\hbox{$<$}\;}

\begin{document}

\title{Toward Eclipse Mapping of Hot Jupiters}

\author{Emily Rauscher\altaffilmark{1}, Kristen Menou\altaffilmark{1},
Sara Seager\altaffilmark{2,3}, Drake Deming\altaffilmark{4}, James
Y-K. Cho\altaffilmark{5} \& Bradley M.S. Hansen\altaffilmark{6}}

\altaffiltext{1}{Department of Astronomy, Columbia University, 1328
Pupin Hall, 550 West 120th Street, New York, NY 10027}

\altaffiltext{2}{Department of Terrestrial Magnetism, Carnegie
Institution of Washington, 5241 Broad Branch Rd. NW, Washington, DC
20015}

\altaffiltext{3}{Dept. of Earth, Atmospheric, and Planetary sciences,
Massachusetts Institute of Technology, 54-1626, 77 Massachusetts
Ave., Cambridge, MA, 02139}

\altaffiltext{4}{NASA/Goddard Space Flight Center, Planetary
Systems Laboratory, Code 693, Greenbelt, MD 20771}

\altaffiltext{5}{Astronomy Unit, School of Mathematical Sciences,
Queen Mary, University of London, Mile End Road, London E1 4NS, UK}

\altaffiltext{6}{Department of Physics and Astronomy and Institute for
Geophysics and Planetary Physics, University of California, 475
Portola Plaza, Box 951547, Los Angeles, CA 90095, USA}

\begin{abstract}
Recent Spitzer infrared measurements of hot Jupiter eclipses suggest
that eclipse mapping techniques could be used to spatially resolve the
day-side photospheric emission of these planets using partial
occultations. As a first step in this direction, we simulate
ingress/egress lightcurves for three bright eclipsing hot Jupiters and
evaluate the degree to which parameterized photospheric emission
models can be distinguished from each other with repeated, noisy
eclipse measurements. We find that the photometric accuracy of Spitzer
is insufficient to use this tool effectively. On the other hand, the
level of photospheric details that could be probed with a few JWST
eclipse measurements could greatly inform hot Jupiter atmospheric
modeling efforts. A JWST program focused on non-parametric eclipse map
inversions for hot Jupiters should be actively considered.
\end{abstract}


\section{Introduction}

Eclipses can be used as powerful tools to spatially resolve the
photospheric emission properties of astronomical objects. During
ingress and egress, the partial occultation effectively maps the
photospheric emission region of the object being eclipsed. With
arbitrarily high photon statistics, the (spatial) resolving power
attainable with such an astronomical telescope is unrivaled. In the
past, eclipse mapping methods have thus been used with great success
to measure limb darkening laws on stellar companions
\citep[e.g.][]{Warner1971}, to map the surface appearance of Pluto
\citep[see review by][]{Stern1992}, or to confirm the Keplerian nature
of accretion disks in Cataclysmic Variables, on the basis of their
radial emission profiles \citep[e.g.][]{rutten96}. Given the recent
Spitzer measurements of infrared eclipses for three hot Jupiters
\citep{Deming2005, Deming2006, Charbonneau2005}, it is natural to
explore the potential for eclipse mapping of hot Jupiters {
\citep{Williams2006}.}

The regime of atmospheric circulation present on hot Jupiters may be
unlike any of the familiar cases encountered in the Solar System.  Hot
Jupiters are massive, gaseous giant planets found in close, circular
orbits around their parent stars, with orbital periods on the order of
a few days \citep[e.g.][]{Butler06}.  General arguments suggest that
these planets are tidally locked \citep{Rasio1996, Lubow1997,
Ogilvie2004}, so that their permanent day-sides are continuously
subjected to intense stellar irradiation, while their night-sides may
be heated only by a much more modest internal energy flux.  In the
presence of such an uneven energetic forcing, leading to significant
(horizontal) temperature gradients, atmospheric winds will attempt to
redistribute the heat more evenly around the planet. The efficiency of
this horizontal heat redistribution is an important open question for
hot Jupiters as it largely determines their observational properties.
Several attempts have been made to address this general circulation
problem for hot Jupiter atmospheres, using different approaches
\citep[][see also Burkert et al. 2005]{ShowGui,Cho,Menou03,C&S}, and
interesting constraints are now being placed on these models through
the recent infrared measurements.

Three hot Jupiters have been directly detected in infrared with the
Spitzer Space Telescope: HD189733b \citep{Deming2006}, HD209458b
\citep{Deming2005}, and TrES-1 \citep{Charbonneau2005}.  The planetary
day-side flux is deduced from the secondary eclipse depth, when the
planet is hidden behind its star.  Based on such flux measurements at
different wavelengths, several groups have computed spectral emission
models for various prescribed levels of heat redistribution and
concluded that at least some amount of heat redistribution must be
present
\citep{Seager2005,Burrows2005,Fortney2005,Barman2005,Burrows2006,Fortney2006}. These
secondary eclipse constraints remain weak at the moment but they could
rapidly improve in quality in the future.  In addition, a first hot
Jupiter infrared orbital phase curve for the non-eclipsing planet Ups
And~b has been reported very recently \citep{HarrUps06}. The large
inferred amplitude of the phase curve, measuring the day- to
night-side temperature difference, suggests a rather weak role for
heat redistribution in this case. More work is likely needed to
interpret the phase curve reliably, and in particular to address the
complications that may arise from the existence of different regimes
of heat redistribution at different heights in the atmosphere
\citep[e.g.][]{Seager2005,Iro,C&S}.

This recent progress in direct infrared detections of hot Jupiters
leads us to consider the possibility that eclipse mapping techniques
could also be used in these systems {\citep[see
also][]{Williams2006}.} Contrary to existing measurements, which only
constrain disk-integrated quantities, eclipse mapping of a hot Jupiter
can provide a ``spatially resolved map'' of its day-side photospheric
disk, with ``pixel'' shapes and sizes determined by eclipse geometry
and photometric accuracy.  {In principle, one can measure the
emission of any individual ``pixel'' directly from the shape of the
ingress and egress lightcurves, without making any significant
assumption on the nature of the photospheric emission itself.  Such
``non-parametric'' inversion techniques} can be used in the eclipse
mapping context
\citep[e.g.][]{horne85,spruit94,rutten96,rutten98,bobinger00,baptista01},
but in this first study we focus on a much simpler parametric method
to evaluate the accuracy with which details of the planet's day-side
photospheric emission can be constrained with Spitzer now and JWST in
the future.  {Specifically, we use physically-motivated,
parameterized models to describe the photospheric emission properties
of hot Jupiters.  We generate artificial eclipse curves, with
appropriate levels of noise, for the three detected eclipsing hot
Jupiters and we determine how well the various models for their
day-side photospheric emission can be distinguished from each other
with repeated eclipse measurements.  This study is thus strongly
model-dependent.  As will be emphasized throughout this paper, it will
be important in the future to perform non-parametric studies of
eclipse mapping for hot Jupiters, since this would be a much more
reliable way to interpret actual data.}

{\citet{Williams2006} first mentioned the possibility of
resolving the ``surfaces'' of hot Jupiters with secondary eclipses.
These authors simulated secondary eclipse lightcurves and introduced
the concept of a ``uniform time offset'' as a useful observable to
differentiate the lightcurves of various photospheric emission models
from that expected for spatially uniform emission.  In particular,
they find that Spitzer should be able to distinguish between the
emission properties predicted by the circulation model of \citet{C&S}
and the case of spatially uniform emission. Our work shares
similarities with that of \citet{Williams2006} but it also differs
from it in a number of ways, most significantly in that we use the
full shapes of ingress/egress curves as observational diagnostics.}

The set of photospheric emission models considered in our study are
described in Section~\ref{sec:models}.  Section~\ref{sec:method}
explains our method to simulate eclipse curves and generate the
associated observational noise, as well as the statistical tools we
use to analyze the data set resulting from repeated eclipse
simulations. Section~\ref{sec:spitzer_analysis} details the results of
our parameter space survey for Spitzer, while
Section~\ref{sec:jwst_analysis} focuses on JWST. We summarize our main
results and propose extensions of this work in
Section~\ref{sec:summary}.

\section{Photospheric Emission Models} \label{sec:models}

Throughout this analysis, we postulate the existence of a well-defined
photospheric surface for any spectral band of observation, so that the
problem of generating ingress/egress lightcurves is reduced to that of
specifying temperatures and/or emission fluxes at various locations on
the planet's photospheric disk.

At the simplest level, there are two extremes for horizontal heat
redistribution in hot Jupiter atmospheres: complete or none.  In the
case of complete heat redistribution, presumably achieved by
atmospheric wind and wave transport, a nearly constant temperature
across the planetary disk is to be expected (modulo limb darkening
effects). In the absence of any heat redistribution, however, a strong
temperature gradient (set by purely local radiative equilibrium) is
expected on the day-side, while the night-side would remain uniformly
very cold under the influence of the modest internal energy flux.  In
between these two extremes, photospheric disk emission properties
would be set by a complex non-linear combination of radiative and
advective processes in the atmosphere. Two groups have performed
global numerical simulations exploring the circulation regime in hot
Jupiter atmospheres.  The multi-dimensional model presented by
\citet[][see also Showman \& Guillot 2002]{C&S} predicts that winds
could advect the hottest part of the atmosphere $\sim$60\degrees in
longitude downstream of the sub-stellar point.  The turbulent
shallow-layer models presented by \citet{Cho} and \citet{Cho06}
predict the emergence of temporally variable temperature structures
associated with two large-scale polar vortices, revolving around the
planet. The observability of such features, however, was found to
depend on a combination of global wind strength and net radiative
forcing in the atmosphere. In particular, a situation approaching pure
radiative equilibrium can be reached in the limit of weak atmospheric
winds and transport. For a detailed parameter space study, see
\citet{Cho06}.

\begin{figure}[ht]
\begin{center}
\includegraphics[width=0.4\textwidth]{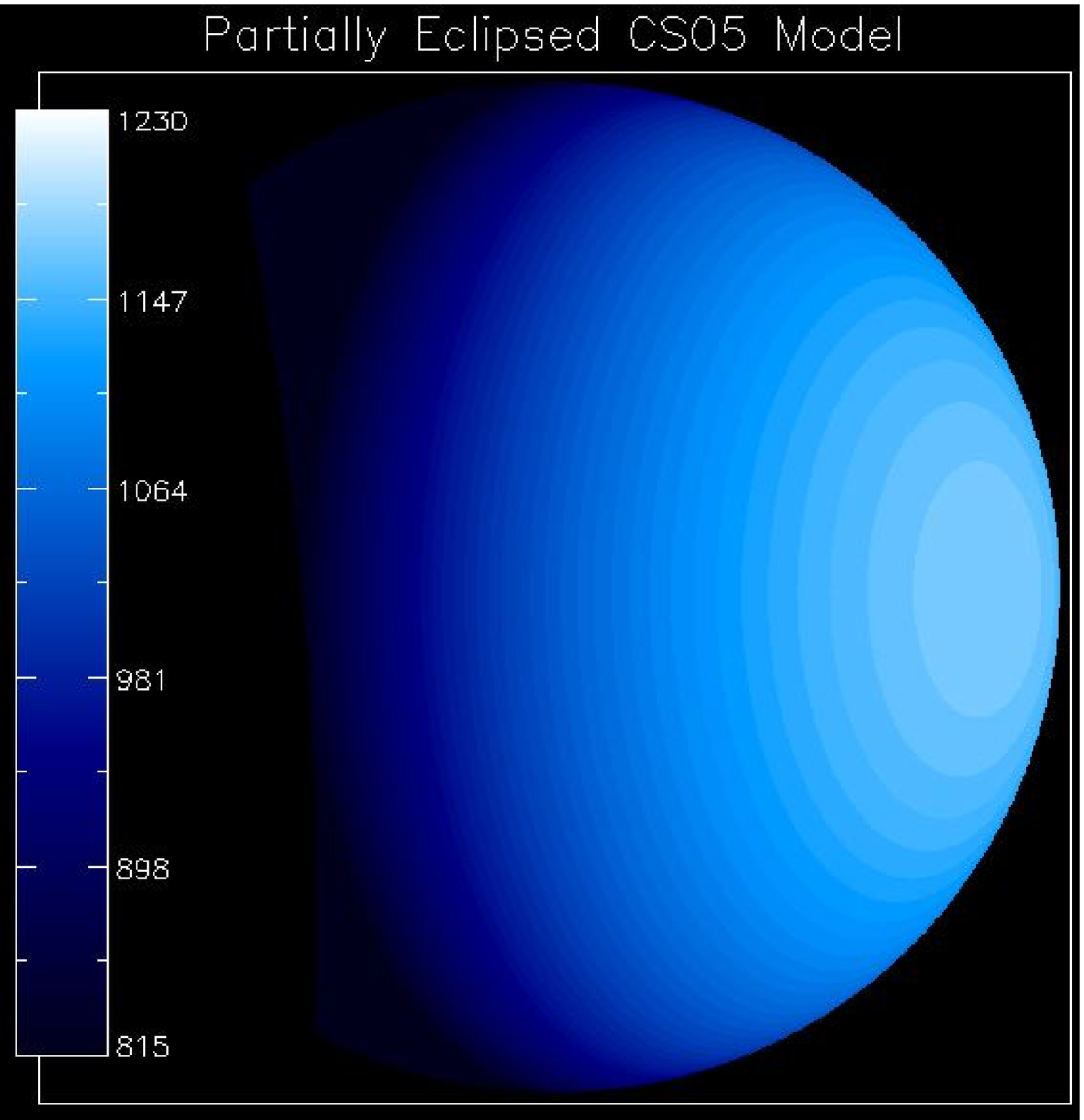}
\includegraphics[width=0.4\textwidth]{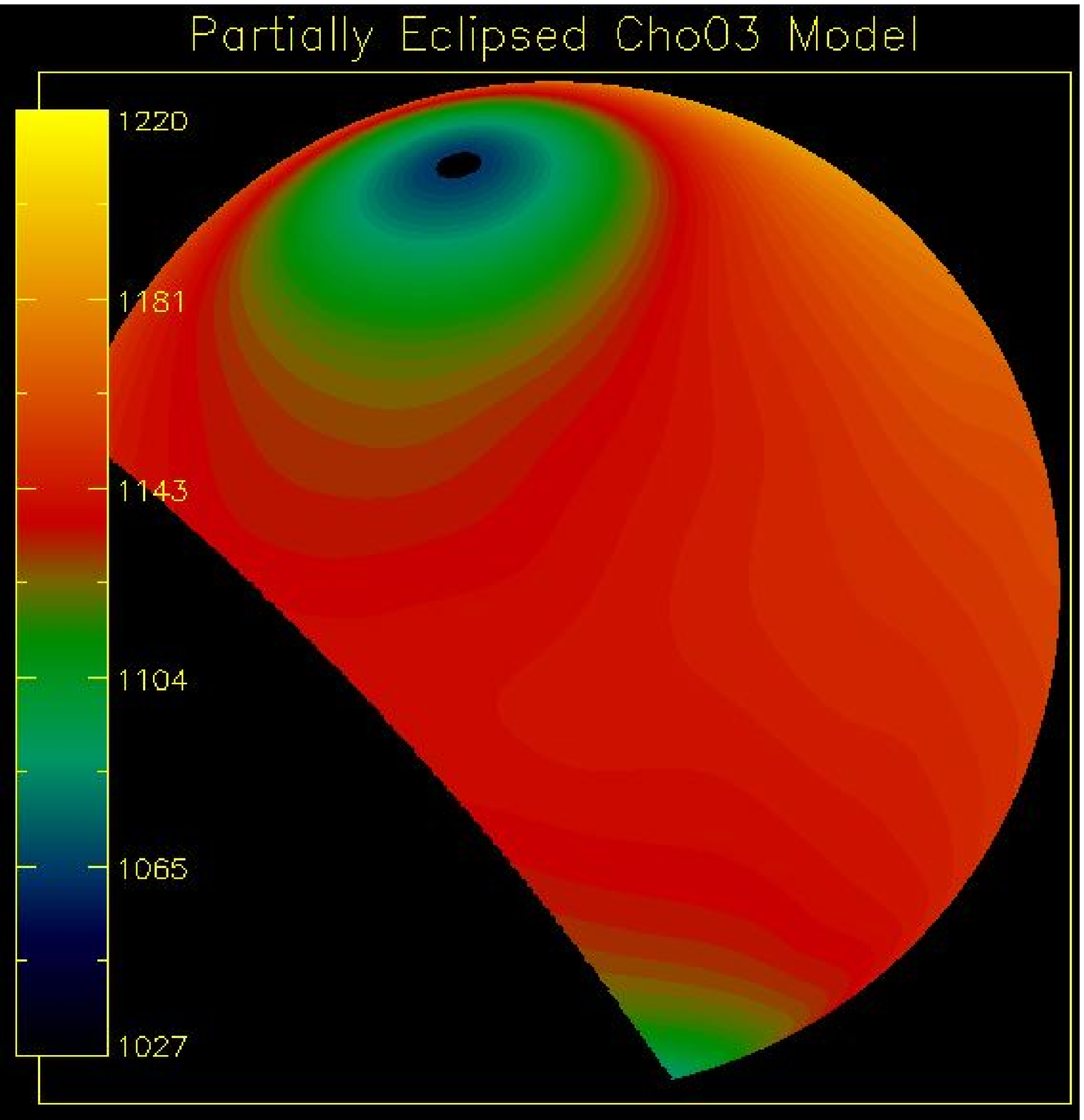}
\caption{Partially eclipsed temperature maps in a CS05-like model of
 TrES-1 (\emph{left}) and a Cho03-like model of HD189733b
(\emph{right}).  The color-temperature scale (in K) is shown on the left of each panel.  Notice how the different system geometries affect the
orientation and shape of the eclipsing stellar limb, and consequently
the detailed shape of the ingress/egress curves.}
\label{fig:prettypics}
\end{center}
\end{figure}

These various scenarios provide us with four distinct and physically
well-motivated models for hot Jupiter day-side photospheric emission, from
which we can build synthetic ingress/egress lightcurves. We label and
parameterize these four scenarios as follows:

\begin{description}

  \item[RadEq Model:] The radiative equilibrium model corresponds to
  the no heat redistribution case, for which one expects a day-side
  temperature profile $T \propto \cos^{1/4}(\theta)$, where $\theta$
  is the angle away from the sub-stellar point. {This arises simply
  from the geometry of stellar irradiation on the day-side.} We assume
  a constant temperature night-side. The specific functional form that
  we adopt for the photospheric temperature is $T(\theta)=\left[ T_0^4
  \cos(\theta) + T_{\rm night}^4 \right]^{1/4}$ for
  $0\leq\theta<90$\degrees and $T=T_{\rm night}$ for
  $\theta\geq90$\degrees (on the night-side).  We calculate the
  radiative equilibrium temperature at the sub-stellar point as
  $T_0=T_*\sqrt{R_*/a}$, where $T_*$ is the effective temperature of
  the host star, $R_*$ is the stellar radius, and $a$ is the orbital
  semi-major axis of the planet.

  \item[Uniform Model:] In this model, complete, efficient heat
  redistribution is assumed, resulting in a constant temperature
  everywhere at the photosphere.  We adopt $T=\left[T_p^4+T_{\rm
  night}^4\right]^{1/4}$. Here the ``redistributed'' equilibrium
  temperature is calculated as $T_p=T_*\sqrt{R_*/2a}\ (=T_0/\sqrt{2})$.

  \item[CS05-like Model:] This is a parameterization of the
  atmospheric model of \citet{C&S}, in which a hot spot is shifted by
  $\sim$60\degrees in longitude downstream of the sub-stellar point.
  For the hot spot itself, we adopt the same functional form as in the
  RadEq model above, except that the hottest region is only $\sim$400K
  hotter than the cold ``night-side'' (and shifted away from the
  sub-stellar point by 60\degrees in longitude). The left panel in
  Figure~\ref{fig:prettypics} shows the corresponding temperature map
  partially eclipsed.

  \item[Cho03-like Model:] This is a snapshot temperature map taken
  from one of the atmospheric models of \citet{Cho} and \citet{Cho06}
  for HD209458b. Two cold spots associated with circumpolar vortices
  rotate around the planet semi-periodically, with a period typically
  several times that of the planet itself. The specific shallow-layer
  model parameters in the simulation used here are: global wind speed
  $\bar{U}=400$~m~s$^{-1}$ and relative amplitude of thermal forcing
  $\eta=0.2$ \citep[for a full discussion of the parameter space
  see][]{Cho06}.  The cold spots are $\sim$10\% cooler than the
  average atmospheric temperature.  A fixed phase of the atmospheric
  temperature pattern has been chosen here for simplicity (it rotates
  around the planet with time), approximately in such a way that the
  greatest difference in ingress/egress lightcurve shape is achieved
  relative to the RadEq model lightcurve shape.  The right panel in
  Figure~\ref{fig:prettypics} shows the corresponding temperature map
  partially eclipsed.
\end{description}

We expect these parameterized, physically-motivated models to cover a
wide enough range of possibilities for us to be able to assess the
level of photospheric details that can be probed via the corresponding
ingress/egress lightcurves. In particular, in the order
RadEq/CS05-like/Cho03-lik3/Uniform, these models constitute a sequence
of increasing deviation from simple radiative equilibrium, towards
photospheric temperature uniformity.

We use these photospheric emission models to simulate ingress/egress
curves for the three, bright hot Jupiters with Spitzer eclipse
measurements: HD189733b, HD209458b, and TrES-1.  All the system
parameters used in our analysis are listed in Table 1. The range of
temperatures in each photospheric model is adjusted to appropriate
values for each planet of interest, using that system's
parameters. From the above description of the RadEq and Uniform model,
it is clear that we have assumed negligibly small albedos in all
cases. We find that a change of Bond albedo from $0$ to $0.3$ \citep[a
range characteristic of most Solar System planets;][]{dePater2001} has
little influence on our main results, considering other systematic
effects present. For computational reasons, we set the value of
$T_{\rm night}$ to $700$~K in all our models. Here again, we find that
this choice is not crucial because very little of the night side is
generally visible to a distant observer, given the tidally-locked
eclipse geometries considered here.

From the detailed model of \citet{C&S}, values of $T_0$ and $T_{\rm
night}$ for HD209458b's photosphere are approximated as 1450K and
1010K, respectively.  In our CS05-like models for the two other hot
Jupiter systems, these values are rescaled in linear proportion to the
value of the ``redistributed'' equilibrium temperature of the planet
of interest (that is $T_p=T_*\sqrt{R_*/2a}$ for zero albedo; see Table
1).  In the detailed HD209458b model of \citet{Cho,Cho06} used here,
the hottest and coldest temperatures reached overall are approximately
1460K and 1230K, with an average temperature of 1390K.  In our
Cho03-like models for the two other hot Jupiter systems, these values
are also rescaled in linear proportion to the value of the
``redistributed'' equilibrium temperature of the planet of interest.

The conversion from local photospheric temperature to emission
contributed by each apparent area element of the planetary disk (see
\S\ref{sec:simulation} for projection details) is performed in two
different ways. In a set of blackbody emission models, a simple Planck
function is used. In a second set of models, a grid of detailed
atmospheric spectral models derived from plane-parallel calculations
is used to evaluate the local emission for effective temperatures in
the range 700-2000K.  No temperatures outside of this range of
validity for the spectral models were allowed in our photospheric
emission models. The plane-parallel spectral models are essentially
extensions of the cloudless models described in
\citet{Seager2005}. Contrary to more detailed calculations
\citep[e.g.][]{Barman2005}, we ignore the 2D geometry of the radiative
transfer problem when we use these plane-parallel calculations and
implicitly assume isotropic emission of each planetary surface element
in our photospheric emission models.

Planetary limb darkening (or brightening) could in principle be
important in hot Jupiter atmospheres. It is not included in our basic
models given the very simple method adopted (which assumes a
well-defined photospheric temperature field and an isotropic
emission).  To estimate the effect that strong limb darkening could
have on our overall analysis, we have included an optional limb
darkening function in our emission models which can be applied to any
of the models under consideration.  The functional form adopted for
this strong limb darkening is chosen to correspond to a geometrical
decrease of the emitted flux with the cosine of the angle away from
the sub-observer point.  (i.e. $F=F_0 \cos(\phi)$, where $\phi$ is the
angle away from the sub-observer point).  Since our analysis relies on
well-defined photospheric temperatures, we have implemented this in
our ``limb-darkened'' models with a simple decrease of the
photospheric temperature as $\cos^{1/4}(\phi)$, superimposed on the
photospheric temperature field of any of the four emission models
defined above.

Another systematic source of errors for hot Jupiter eclipse mapping
studies could be related to finite eclipse timing uncertainties. Even
for planets with measured orbital eccentricities consistent with zero,
the presence of an as yet undetected perturbing companion could lead
to significant timing variations from one eclipse to the next
\citep{Agol2005,HolMur05}. Careful measurements of transit times
before and after any secondary eclipse could be used to put stringent
constraints on the magnitude of these timing uncertainties. A thorough
exploration of timing variation uncertainties for eclipse diagnostics
is beyond the scope of the present analysis.  Here, to evaluate simply
the magnitude of these effects, we consider in some of our models the
effects of arbitrarily time-shifting lightcurves produced by two
different photospheric emission models, in such a way as to bring the
mid-eclipse values closer to each other. This makes efforts to
distinguish the models more difficult, even though we are not formally
fitting the eclipse center time in our analysis. In principle, the
influence of a perturbing planet results in timing offsets which vary
in amplitude from orbit to orbit.  Here we are only considering the
simple consequences of a constant, detrimental temporal shift in our
models.  Although \citet{Agol2005} have shown that a wide range of
timing variations are possible in general {(from seconds to minutes
and more), we will focus here on a representative value of 7 seconds
for definiteness, on the assumption that careful transit monitoring
could plausibly constrain secondary eclipse timing uncertainties to
$\lsim 10$ seconds in any given system \citep[which is about an order
of magnitude improvement over current timing accuracies;
e.g.][]{Deming2006}.}

\section{The Method} \label{sec:method}

It is clear that current Spitzer data for a single eclipse in any of
the three systems studied here do not contain enough information to
determine the shape of the ingress or egress lightcurves with much
accuracy \citep[e.g.][]{Deming2006}. Multiple Spitzer eclipse
observations would be required to derive meaningful constraints on the
day-side photospheric emission properties of these hot Jupiters.  The
ability to distinguish between various photospheric emission models
must then depend on whether or not the number of repeated observations
needed is prohibitively large.  To evaluate this, we simulate eclipse
light curves produced by different photospheric emission models, add
noise in proportion to known measurement errors and statistically
determine the number of observations needed to differentiate between
two specific models. In other words, our analysis provides a
model-dependent estimate of the level of atmospheric details that can
be probed with repeated ingress/egress measurements.

  \subsection{Simulating Eclipses} \label{sec:simulation}

To simulate eclipse light curves, a photospheric temperature model for
the planet is orthographically projected onto a disk, which is then
eclipsed using the known geometry of the system.  At each point in
time during ingress or egress, the emission from non-eclipsed regions
on the planetary disk is integrated, in proportion to the (apparent)
area of each visible disk surface element.  Two analyzes are performed
simultaneously.  One focuses simply on the bolometric emission under
the assumption that each surface element emits like a blackbody at the
corresponding photospheric temperature. The other analysis is spectral:
the contribution to the monochromatic flux at Earth is calculated for
each apparent surface element by using the local photospheric
temperature to interpolate in the above-mentioned grid of
plane-parallel spectral emission models.  Spectral contributions are
then integrated over specific Spitzer response curves, or the
wavelength ranges of the future JWST NIRSpec gratings, to determine the
amount of light that would be collected by these instruments.  The
spectral models used are subject to strong systematic effects in
specific spectral regions \citep[e.g. flux peak at $\sim
4$\micron, as seen Fig~1 of][]{Seager2005}. Testing how strongly our
main results depend on such systematic spectral features is the reason
why we use an alternative, very idealized bolometric blackbody
emission model in some of our eclipse simulations.

A 2D polar grid is constructed to describe the projected planetary
disk, with 1000 radii sampled between 0 and $R_p$ (the planetary
radius) and 1500 polar angles sampled between 0 and 2$\pi$.  The
radial resolution also determines the temporal resolution of our
simulated eclipse curves since the stellar limb moves across the disk
at a constant horizontal rate.  We account for the curvature of the
stellar limb, assuming it to be a perfectly opaque disk.  We have
tested our eclipse simulation apparatus by comparing results to the
analytical solution described in \citet{Warner1971}. With a
[$r$,$\theta$] resolution of [1000,1500], a comparison to the
analytical lightcurve solution for a straight stellar limb gives
agreement at better than the 1\% level (with growing errors towards
the very end/beginning of ingress/egress because of grid
under-sampling).  The same level of accuracy is achieved in the case
of a spherically curved stellar limb.

In our analysis, we account for the influence of each system's
specific geometry (i.e. inclination, ratio of planet and stellar
radii, semi-major axis) on details of how the planetary disk is being
eclipsed.  Table~1 lists the geometrical parameters adopted for the
three systems of interest.  The orbital velocity is calculated as
$v=2\pi a/P$, where $a$ is the planet's orbital semi-major axis and
$P$ is the corresponding orbital period.  The vertical offset (from
the observer's point of view) between the centers of the planetary and
stellar disks, $h$, is calculated as $h=a \sin(90^\mathrm{o}-i)$,
where $i$ is the orbital inclination.  The duration of ingress/egress
is $t_{\rm gress}=(\sqrt{(R_*+R_p)^2-h^2}-\sqrt{(R_*-R_p)^2-h^2})/v$,
where $R_*$ is the stellar radius.  Figure~\ref{fig:prettypics}
illustrates the eclipsing geometries for the TrES-1 and HD189733
systems.  For simplicity, we ignore the effects of the small planetary
rotational offset occurring between ingress and egress and consider
instead that the sub-stellar point remains directly aligned towards
the observer in all the models.  The actual shift between ingress and
egress is $\sim 10$\degrees for the three systems of interest. We have
performed explicit tests and found that such a small rotational offset
of the photospheric emission model generally has little effect on our
main results, as compared to other more important sources of
uncertainty.

\begin{figure}[ht!]
\begin{center}
\includegraphics[width=\textwidth]{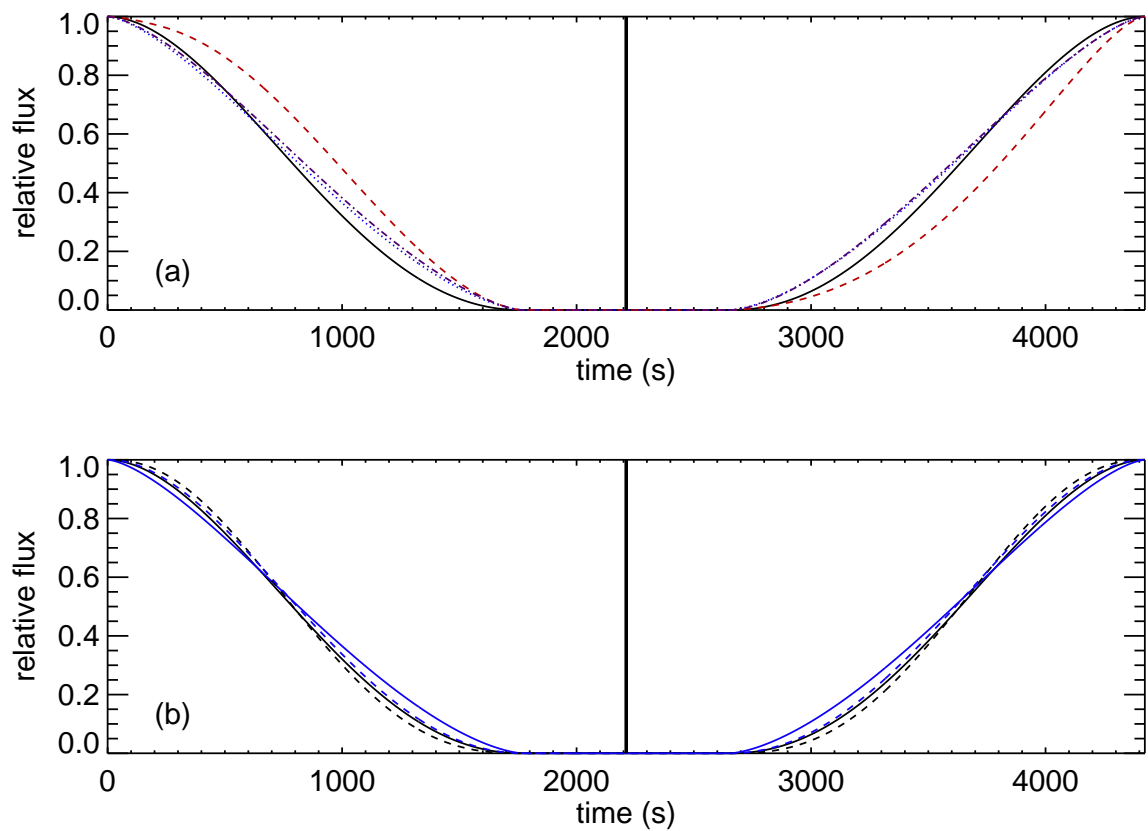}
\caption{Simulated ingress and egress curves for different
photospheric emission models of HD189733b: (a) RadEq model
(\emph{black}), Uniform model (\emph{dotted blue}), CS05-like model
(\emph{dashed red}), and Cho03-like model (\emph{dot-dashed purple});
(b) RadEq model (\emph{black}), Uniform model (\emph{blue}), RadEq
model with strong limb darkening (\emph{dashed black}), and Uniform
model with strong limb darkening (\emph{dashed blue}).  Our analysis
considers only the ingress and egress phases (separated by the
vertical line) and thus excludes the full occultation period.}
\label{fig:lightcurves}
\end{center}
\end{figure}

As an example, Figure~\ref{fig:lightcurves} shows the bolometric
blackbody ingress and egress light curves obtained in the case of
HD189733b for each of the photospheric emission models considered in
our analysis. Dashed lines highlight the effects that strong planetary
limb darkening has on the RadEq and Uniform models.  It is obvious
from these curves that the differences between the various
photospheric emission models under consideration are subtle.  It is
also important to realize the existence of some level of degeneracy in
our models. For example, during egress (only), there is little
difference between the curves predicted by the Cho03-like and the
Uniform models because the stellar limb exposes the Cho03-like
photospheric emission features in such a way that contributions from
hot and cold regions more or less cancel out. {Finally, let us
emphasize that the various photospheric emission models introduced
above only define the shapes of ingress/egress lightcurves, in units
of relative flux, in our modeling strategy. An absolute flux unit
common to all these models is effectively selected only when a noise
magnitude is chosen to generate a system- and instrument-specific
stream of simulated data, as we shall now describe.}

  \subsection{Statistical Analysis} \label{sec:stats}

With simulated ingress/egress light curves in hand, we can use known
or estimated measurement errors to simulate noisy eclipse data. From
these simulations, we can statistically determine how much data is
needed to differentiate between two specific photospheric emission
models.  For ease of comparison and definiteness, we chose the model
without any heat redistribution, that is the RadEq model, to be our
reference model. We generate an incrementally increasing amount of
simulated eclipse data for this model and determine when it is that
model lightcurves produced for other photospheric emission models
become inconsistent with the accumulated stream of data (i.e. as a
function of the number of repeated eclipse measurements).  Our choice
of the RadEq model as a reference is {arbitrary, in the sense that any
other model could be used as a reference for our statistical
analysis.} Notice, however, that by ruling out another model against
the RadEq one, one effectively rules out some level of heat
redistribution in the atmosphere studied. While any combination of two
photospheric emission models could in principle be compared against
each other, the interest of doing so is limited by the fact that all
comparisons remain necessarily model-dependent. In the end, with our
analysis we are only trying to assess the level of photospheric detail
that could be distinguished by actual measurements.  We will later
suggest that non-parametric methods be considered for the study of hot
Jupiter eclipses in the future.

\begin{figure}[ht!]
\begin{center}
\includegraphics[width=0.7\textwidth]{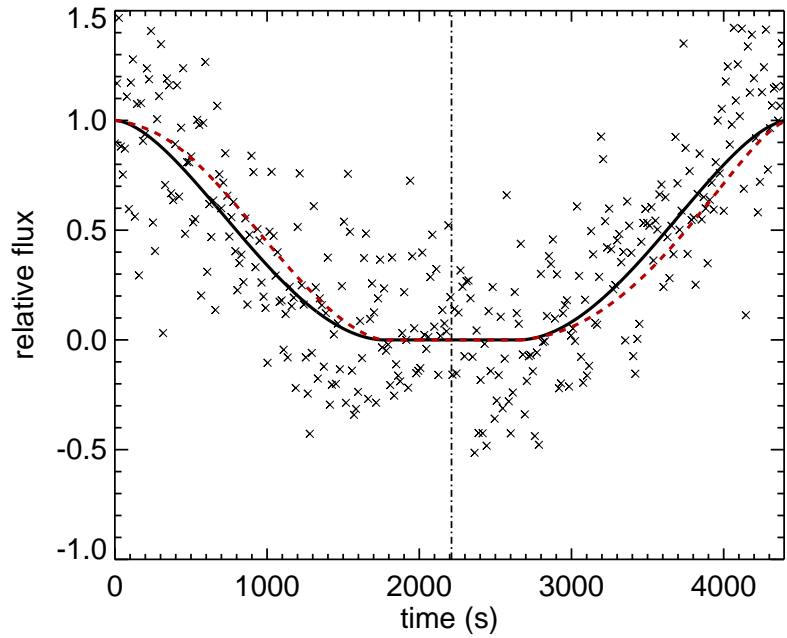}
\caption{Example of ingress/egress simulated data for HD189733b in
Spitzer's 8\micron~IRAC band, for single-measurement noise of variance
$\sigma_1=0.31$ (crosses) {based on existing Spitzer measurements.}
For comparison, two noise-free curves are shown for the RadEq model
(\emph{black}) and CS05-like model (\emph{red dashed}). Here again,
our analysis excludes the full occultation period.}
\label{fig:simdata}
\end{center}
\end{figure}

Throughout this analysis we work in units of relative flux. The
planetary flux is normalized to its full day-side value, i.e. the
value right before entering secondary eclipse (neglecting small
rotational effects).  A simulated data stream is generated on the
basis of a cadence of observation and a Gaussian noise
single-measurement error of variance $\sigma_1$ (in units of relative
flux). The values of $\sigma_1$ adopted, as well as the cadence used
to calculate the number of measurements occurring during one ingress
or egress period, are estimated from previous measurements or simple
extrapolations, as described in Section~\ref{sec:spitzer_errors} for
Spitzer, and Section~\ref{sec:jwst_errors} for
JWST. Figure~\ref{fig:simdata} shows an example of such a simulated
data stream for HD189733b in one of Spitzer's IRAC bands. Noiseless
ingress/egress curves for two photospheric emission models (including
the RadEq one used to generate the simulated noisy data) are
over-plotted for comparison.  Again, it is clear that repeated
eclipses will be needed in this case if one wishes to discriminate
between the two models, given with the level of noise in the data.

The quality of any photospheric emission model in fitting a set of
simulated data is evaluated by calculating the reduced $\chi^2$ value
for that model compared to the data.  Since we are creating the
simulated data from the reference RadEq model, with a sufficient
number of repeated eclipse measurements, any other emission model will
eventually provide a poor fit to the data (as measured by the reduced
$\chi^2$ value).  The process of simulating data is repeated, as
needed, for an increasing number of eclipses, $N$, with identical
error properties but a random temporal offset added to each new
eclipse set.  This offset is added to account for the likely lack of
accurate phasing even between successive eclipses in a real data
set. For each value $N$ of the number of accumulated eclipses, reduced
$\chi^2$ values are calculated comparing any photospheric emission
model of interest to the data until, after enough measurements, models
(different from the RadEq one) are ruled out at the 99\% confidence
level by the accumulated data set.

\begin{figure}[ht!]
\begin{center}
\includegraphics[width=0.7\textwidth]{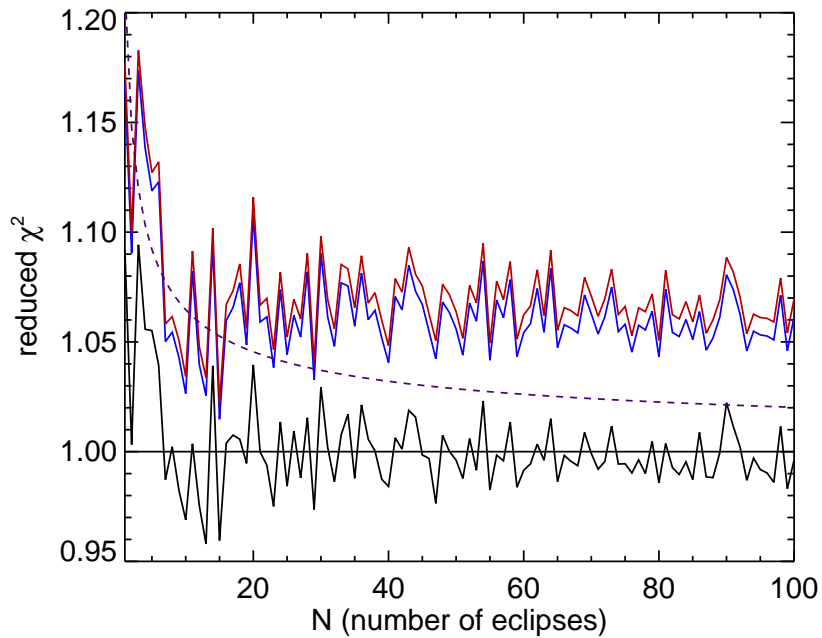}
\caption{Reduced $\chi^2$ values of three different model fits to a
simulated dataset for HD189733b, as a function of the number of
repeated eclipse measurements, $N$. The single-measurement noise
variance adopted in the simulated ingress/egress dataset is $\sigma_1=
0.31$.  The 99\% confidence limit above which models are ruled out is
shown as a dashed line.  The RadEq model used to generate the
simulated ingress/egress dataset is shown as a \emph{black} line. With
a large enough set of eclipse data, other models (\emph{red line}:
CS05-like model; \emph{blue line}: CS05-like model with detrimental
7~second timing offset) are eventually ruled out at high
significance.}
\label{fig:chitest}
\end{center}
\end{figure}

Figure~\ref{fig:chitest} illustrates this procedure.  At each $N$, all
models of interest are compared to the same set of simulated data, so
that the random fluctuations in the data set affect all of the reduced
$\chi^2$ values in a similar way, as is apparent in the fluctuation
pattern common to each reduced $\chi^2$ curve.  The value of $N$ at
which a reduced $\chi^2$ line crosses the 99\% confidence level and
does not return below it, $N_0$, is our measure of the number of
eclipses needed to reliably differentiate between that model and the
reference RadEq model.\footnote{Alternatively, one could use the
number of eclipses for which a reduced $\chi^2$ line first crosses the
99\% confidence level to declare the corresponding model unfit. We
have found that this less conservative criterion is unreliable.
Figure~\ref{fig:gt105} illustrates well how it fails.} For example,
Figure~\ref{fig:chitest} shows that the RadEq model can be
differentiated from the CS05-like model with approximately 15
eclipses, since the latter is then ruled out according to our
criterion.  When a detrimental 7-second timing offset is added in the
CS05-like model (in such a way that it makes harder to differentiate
against the RadEq model), a total of $\sim$30 eclipses is required
according to our criterion.  A close inspection of
Figure~\ref{fig:chitest} suggests that this substantial difference
could simply be a statistical fluke resulting from the use of one
statistical realization of the data stream.

\begin{figure}[ht!]
\begin{center}
\includegraphics[width=0.7\textwidth]{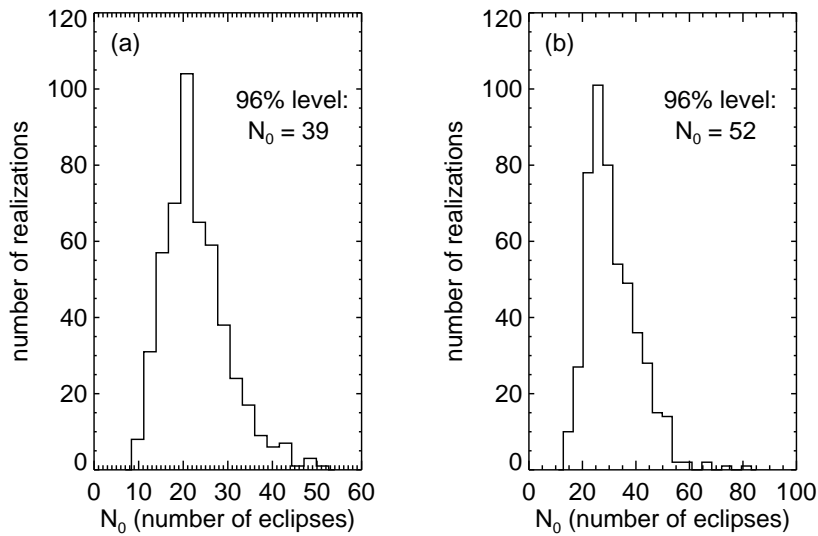}
\caption{Histograms of $N_0$, the number of eclipses needed to
differentiate at the $99\%$ confidence level between fits to the
correct underlying RadEq model and a CS05-like model (a), or a
CS05-like model with an imposed 7 second detrimental timing offset (b).  These
distributions emerge from 500 independent realizations of repeated
eclipse sequences like the one shown in Figure~\ref{fig:chitest}.
$N_0$ is defined as the value of $N$ in such sequences at which each
model's reduced $\chi^2$ curve reliably crosses the 99\% confidence level.}
\label{fig:hists}
\end{center}
\end{figure}

\begin{figure}[ht]
\begin{center}
\includegraphics[width=0.6\textwidth]{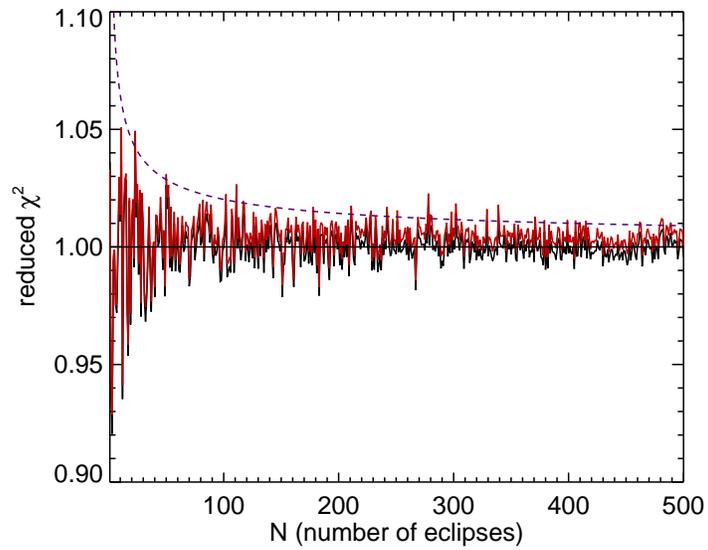}
\caption{Example of the reduced $\chi^2$ values as a function of the
number of eclipses, $N$, for data compared to the RadEq model
(\emph{black}) and the Uniform model (\emph{red}) for HD189773b using
Spitzer's 8\micron~IRAC band.  The models cannot be differentiated by
this method.} \label{fig:gt105}
\end{center}
\end{figure}

To account for statistical variations in the value of $N_0$ derived
from single realizations of the data stream, we repeat the above
procedure 500 times (i.e. we generate 500 data streams with identical
parameters but different random number generators). From these 500
realizations, we obtain a representative distribution of $N_0$ values
for a fixed set of model parameters.  Figure~\ref{fig:hists} shows
examples of such distributions of $N_0$ values {for Spitzer
measurements}.  From these distributions, we now see clearly that a
7-second detrimental offset in timing applied to the CS05-like model
does indeed make it harder to differentiate that model from the
reference RadEq model.  To be able to distinguish the fiducial RadEq
model from the CS05-like model in 96\% of the data stream
realizations, $39$ eclipses are needed. This number jumps to $52$
eclipses if the detrimental 7-second timing offset is imposed on the
CS05-like model curve, thus showing that eclipse timing uncertainties
do affect the eclipse diagnostics being discussed here

We note that other parametric methods, based on different eclipse
diagnostics, exist.  In particular, \citet{Williams2006} discuss the
overall time shift between a Uniform model and an implementation of
photospheric emission for the detailed atmospheric model of
\citet{C&S}.  One advantage of our method over the time shift
diagnostics is that we use more information, that is the full shape of
the ingress/egress curves, which makes it less sensitive to
degeneracies.  To illustrate this point, we note that a comparable
mid-eclipse time shift deviation from the prediction of the CS05-like
model is expected for both the RadEq and the Uniform models, making it
difficult to discriminate between these two models solely on the basis
of a time shift (see Fig.~\ref{fig:lightcurves}).  An adequately
phased/rotated Cho03-like model could also induce a similar time shift
relative to the RadEq and Uniform model predictions. Clearly, the
issue of degeneracies is inherent to all parametric methods and we
have already illustrated how our models also suffer from this
limitation. In the future, non-parametric inversion techniques may
thus be the best way to address this difficulty.

  \subsection{Estimating Spitzer Errors} \label{sec:spitzer_errors}

So far, we have considered in detail only the specific case of
HD189733b observed by Spitzer at 8\micron~with IRAC. The parameters
used to produce the data stream in this case were inferred from
actual observations of that system with Spitzer and from observations
of another system with IRAC at 8\micron. Specifically, the
single-measurement error was taken to be $\sigma_1=0.31$ and 134
measurements were assumed to be collected per ingress/egress period.
Here we describe how we estimate these two numbers for every possible
combination of instrument and planetary system of interest.  This
necessarily involves some level of extrapolation since many such
observations have not been performed (or reported) to date.

In the previous section, we defined $\sigma_1$ as the error per
observational point, for unity eclipse depth (i.e. in our relative
flux units).  In absolute units, this corresponds to the
single-measurement error divided by the eclipse depth flux-equivalent,
which can both easily be obtained from previous Spitzer measurements
when they exist.  We have compiled the results of \citet{Deming2005} using
MIPS on HD209458b, \citet{Charbonneau2005} using the 4.5 and
8\micron~IRAC bands on TrES-1 and \citet{Deming2006} using IRS on
HD189733b.  The values of $\sigma_1$ derived from these measurements
are shown in bold in Table~2.

To estimate the errors for every possible combination of system and
instrument, we pick an instrument, a specific spectral band, and
rescale a known measurement error to derive the corresponding value for
another system.  Assuming that simple Gaussian noise statistics
and the bright source limit are
 applicable, so that $\sigma_F=\sqrt{F}$, we expect the errors
between system A and system B to scale approximately with the ratios
of system fluxes as:
\[
\frac{\sigma_{1,B}}{\sigma_{1,A}} \simeq
\sqrt{\frac{F_{*,B}}{F_{*,A}}}\left[\frac{F_{p,A}}{F_{p,B}}\right]
\]
\noindent where the $*$ and $p$ subscripts refer to the star and the
planet, respectively.  Using a simple blackbody-like scaling for the
relative stellar and planetary fluxes, and isotropic emission
in all cases, this expression becomes:
\begin{equation}
\frac{\sigma_{1,B}}{\sigma_{1,A}} \simeq \frac{d_B}{d_A}
\left(\frac{R_{*,B}}{R_{*,A}}\right)
\left(\frac{R_{p,A}}{R_{p,B}}\right)^2
\sqrt{\frac{B_\lambda(T_{*,B})}{B_\lambda(T_{*,A})}}
\left[\frac{B_\lambda(T_{p,A})}{B_\lambda(T_{p,B})}\right]
\end{equation}
\noindent where $d$ is the distance to the system, $R$ is the object's
radius, $B_\lambda$ is the Planck function, $T$ is temperature, and
$\lambda$ is taken to be the central wavelength of the instrumental
band under consideration.  The results of this extrapolation for
$\sigma_1$ values are listed in Table~2.

Table 2 also lists in each case the number of single data point
measurements per ingress/egress period.  These values are calculated
from the duration of ingress/egress for each planet, combined with the
cadence of each instrument.  The ingress/egress durations,
$t_{\rm gress}$, are listed in Table 1.  The cadences for existing
measurements are 12.3 s for MIPS, 13.2 s for IRAC, and 14.7 s for IRS.
New observing plans could result in different cadences, but these
values should be sufficiently accurate at the level of detail of our
analysis. We estimate that uncertainties in extrapolated $\sigma_1$
values may affect our results much more significantly.

\section{Results for Spitzer} \label{sec:spitzer_analysis}

We performed an extensive survey of the parameter space of our eclipse models
for each planetary system and Spitzer instrument of interest. The
results of this exploration are summarized in Table~3.

Figure~\ref{fig:gt105} shows a typical reduced $\chi^2$ test for what
we estimate to be the best scenario for Spitzer, differentiating between the
RadEq and Uniform models at 8\micron~with IRAC in
the HD189733 system.  Clearly, Spitzer is not sensitive enough to
differentiate between these two models in any practical
terms.\footnote[5]{According to our study, the only possible way to
reliably discriminate between the RadEq model and the Uniform model with
Spitzer, in less than a dozen eclipses or so, would be to discover a new
system at least 10 times brighter than HD189733.}  

It is also impossible for Spitzer to differentiate the Cho03-like
model from the RadEq model. One additional difficulty with the
Cho03-like model is that the rotation of the temperature pattern would
cause the detailed shape of the ingress/egress curves to change
significantly over time if too many successive eclipses are required
to reach good photometric accuracy.  An instrument sensitive enough to
differentiate the Cho03-like model from the RadEq model would probably
have to do so in no more than a few successive eclipses to avoid
excessive variations of ingress/egress shapes.  In the specific
Cho03-like model used here, the circumpolar vortices are $\sim$10\%
cooler than the average photospheric temperature. At the level of
detail probed by Spitzer, we find that this model is indistinguishable
from the Uniform model.  \citep[However, see][for more extreme
versions of this class of models.]{Cho06}

Of all the models we have considered in our survey, the only one that
Spitzer could possibly distinguish from the RadEq reference model is
the CS05-like model, in the bright HD189733 system. Even in this case,
from 9 to 52 eclipses are required (see Table 3 and Figure 4). This
would be a very expensive observing program considering the strongly
model-dependent nature of the diagnostics involved.

If, instead of the spectral predictions of plane-parallel atmospheric
calculations, we use simple, bolometric, blackbody photospheric
emission models, we find in general that fewer eclipses are necessary
to permit discrimination between two models (see comparisons in
Table~3). Note that this difference is not due to the more efficient
character of blackbody emission since all our ingress/egress models
are eclipse-depth normalized in such a way that it is the normalized
ingress/egress shape that is being used as a diagnostic. Rather, this
shows that details of how the spectral energy distribution varies with
local photospheric temperature do affect the diagnostics of our
analysis.  As additional observational constraints on hot Jupiter
atmospheres become available, some classes of emission models
(e.g. with or without clouds) will be favored over
others. Nevertheless, the sensitivity of eclipse diagnostics to
details of the spectral emission model assumed could remain
significant. This is another important reason to favor diagnostics
based on non-parametric inversion techniques in the future.

While limb darkening could worsen the $N_0$ statistics discussed so
far, by reducing the information content of photospheric emission
regions close to the planetary limb, we have not systematically
explored this possibility in the context of Spitzer observations given
the already poor statistics. We come back to this issue in the context
of JWST observations below.

\section{The Case For JWST} \label{sec:jwst_analysis}

Since Spitzer will likely not be able to constrain photospheric
temperatures and atmospheric models at a very meaningful level
{with eclipse mapping,} we turn our attention to the upcoming
James Webb Space Telescope (JWST). The much increased sensitivity of
JWST results from a combination of larger collecting area and reduced
background. Of the instruments currently planned for JWST, we chose to
focus on NIRSpec for our comparative analysis.  This instrument will
have three spectral gratings, covering the range of wavelengths from
1-5\micron.  Spectra are not required for our analysis (or for eclipse
mapping in general), but the dispersion of light provided by the
gratings turns out to be useful in avoiding possible instrument
saturation for the three bright systems of interest here.  It may be
possible to use the NIRCam instrument to study eclipses in dimmer
systems than the three bright ones considered here, thus increasing
the total number of potential target systems, but we have not explored
this possibility in detail. Our goal here is a first, very preliminary
assessment of JWST capabilities in terms of eclipse mapping, relative
to the Spitzer case. We caution that it is difficult to estimate
high-contrast measurement errors for an instrument that was not
designed for that task.

\subsection{Estimating JWST Measurement Errors} \label{sec:jwst_errors}

As the JWST design is not yet finalized, our estimate of measurement
errors for NIRSpec is only approximate.  We treat each NIRSpec
spectral grating as a single bandpass, using the spectrally integrated
light only.  We base our estimates on the detailed analysis of
\citet{Valenti2005} for a NIRSpec simulated (spectral) measurement of
a secondary eclipse of HD209458b.  The three gratings G140H, G235H,
and G395H cover the wavelength ranges 1-1.8\micron, 1.7-3\micron, and
2.9-5\micron, respectively.  For HD209458b, the respective exposure
times are estimated to be 2.4, 2.4, and 3.6~s.  Adding a conservative estimate
of 0.9 s for reset between exposures results in cadences of 3.3 and
4.5 s for G140H/G235H and G395H, respectively \citep{Valenti2005}.

A calculation of the single-measurement error, $\sigma_1$, entering our
ingress/egress simulations is more complex and requires estimates of
the number of stellar and planetary photons that would be detected by
each grating during one exposure.  \citet{Valenti2005} simulate noisy
spectra for the HD209458 system, for two hours of exposure time.  To
estimate the eclipse depth in any given grating, we divide the
planetary spectrum by the stellar spectrum and sum over the wavelength
range of that grating.  Since the spectra are summed over two hours of
exposure time, the number of photons collected during one exposure is
easily obtained from the number of single exposures in a two-hour
period. We assume that the noise scales simply with the inverse square
root of the number of detected stellar photons
\citep[like][]{Valenti2005}.  The single-measurement error is then
given by $\sigma_1=(\sqrt{N_{*} \times N_{\rm exp}})/(N_{p})$, where
$N_*$ is the number of stellar photons, $N_p$ is the number of
planetary photons and $N_{\rm exp}$ is the number of unit exposures.
Using Figures 3 and 4 from \citet{Valenti2005}, a spectral resolution
$R=2700$ and a value of $N_{\rm exp}$ based on the cadences listed
above, we deduce the approximate single-measurement errors for
HD209458b listed in italics in Table 2.  Using the same
system-to-system scaling method as before
(\S~\ref{sec:spitzer_errors}), we then extrapolate single-measurement
$\sigma_1$ errors for HD189733b and TrES-1 (see Table 2).

As a consistency check on the $\sigma_1$ values thus derived from
\citet{Valenti2005}, the following simple arguments can also be used
to scale values from the 4.5\micron~Spitzer-IRAC band to the
4\micron~JWST spectral gratings.  Wavelength ranges for these two
bands overlap, but they are not equivalent: the NIRSpec grating covers
roughly 3-5\micron, while the IRAC band covers approximately
4-5\micron.  The additional 3-4\micron~sub-range will allow JWST
gratings to collect more photons and probe a region where the planet
to star flux ratio is somewhat higher \citep[due to the presence of a
flux peak at $\lsim$4\micron; see Fig.~1
of][]{Seager2005}. Assuming that NIRSpec has 1000-fold increased
sensitivity relative to IRAC, that the NIRSpec cadence is about three
times that of IRAC, and that the JWST to Spitzer eclipse depth ratio
in these bands is about a factor of two (from the 4\micron~flux peak),
we estimate a Spitzer-to-JWST $\sigma_1$ ratio of $\sim$50.  This is
consistent with the ratio $58$ listed in Table~2.  While this crude
comparison lends support to the $\sigma_1$ values adopted below in our
JWST analysis, there is clearly a need for better evaluations of the
photometric accuracy expected with JWST for hot Jupiter eclipse
mapping programs.

\subsection{Results For JWST}

The results of our survey of JWST/NIRSpec capabilities are summarized
in Table~3.  The much increased sensitivity allows relatively small
photospheric emission features to be distinguished in the shapes of
ingress/egress lightcurves in just one or a few eclipses.  For
HD189733b and HD209458b, the RadEq model would be easily distinguished
from the Uniform and Cho03-like models by using the 4\micron~grating.
HD189733b is bright enough that any of the gratings would in fact be
very useful, while for HD209458b, the 4\micron~grating is a strongly
preferable choice.  The errors for this grating are expected to be
significantly lower than the errors for the other two, partly because
the planet to star flux ratio is higher at these wavelengths
(according to the similar spectral models we and Valenti et al. 2005
used).  A similar attempt to discriminate between the RadEq model and
the Uniform model may require a prohibitively large number of eclipses
for TrES-1.  With the same 4\micron~grating, our results indicate that
the CS05-like model can be differentiated from the RadEq one on all
three planets. We also find this time that, given the much increased
sensitivity of JWST, applying a detrimental 7~s timing offset to the
CS05-like model has little consequence for attempts to differentiate
that model from the RadEq one.

When we superimpose a strong limb darkening law on our Uniform
photospheric emission model, it can be more difficult to differentiate
that model from the RadEq model. As an example, observing HD189733b
with the 4\micron~grating, one eclipse is still enough to
differentiate between the RadEq model and the limb-darkened Uniform
model, but five eclipses are required to do so with the
1.4\micron~grating.  At the level of precision achievable with JWST,
it will thus be important to self-consistently account for limb
darkening effects for reliable eclipse diagnostics. This is yet one
more reason to focus on non-parametric inversion techniques which, by
construction, automatically incorporate any photospheric limb
darkening present.

As in Section~\ref{sec:spitzer_analysis}, we find that the simplified
bolometric blackbody analysis generally leads to a reduction in the
number of eclipses needed for model differentiation, relative to the
analysis based on the detailed spectral emission models. We interpret
this as indicating that emission in the spectral models vary generally
less with temperature (in relevant narrow spectral bands) than the
Planck function does bolometrically.  We note that a counterexample of
this general trend is found at the shortest wavelengths considered
(the 1.4\micron~grating). For this grating, the range of photospheric
temperatures on HD189733b conspires to make the emission properties in
the spectral model vary more over ingress/egress than in the
corresponding blackbody model.  This time, a lower number of eclipses
is needed to differentiate between the RadEq model and the Uniform
model, according to spectral models.  Altogether, comparisons between
spectral and bolometric blackbody versions of our eclipse analysis,
and their qualitative agreement, suggest that our main conclusions (in
terms of the number of eclipses required for model differentiations)
are not too strongly affected by systematic effects or uncertainties
in the specific cloudless spectral emission models adopted throughout
our analysis.

As a special case, we have also considered the possibility of
differentiating between the Cho03-like and the Uniform photospheric
emission models on HD189733b. We find that the 4\micron~NIRSpec
grating should perform measurements precise enough to make the
distinction between these two photospheric emission models possible
with a single eclipse measurement (see the last three lines in
Table~3). This would alleviate complications arising from temporal
variations of the rotating atmospheric pattern, which are expected in
Cho03-like atmospheric models.  Let us recall that we have fixed the
phase of the weather pattern in all our Cho03-like models to produce
the greatest difference in ingress/egress shapes with respect to the
RadEq model predictions. We selected a phase for the cold circumpolar
vortices that would most easily be distinguished from that RadEq
model.  In a realistic situation, this specific phase would only be
observed by chance coincidence. Our goal here was merely to determine
the level of photospheric details that can be probed with JWST.  This
exercise shows that features of amplitudes and sizes similar to those
of the circumpolar spots present in the Cho03-like model considered
here (approximately 10\% cooler than the average planetary
photospheric temperature) are within the reach of JWST.

\section{Summary and Conclusions} \label{sec:summary}

{ The shape of ingress and egress curves produced when a hot Jupiter
is gradually eclipsed by its parent star contains detailed information
on the planet's photospheric emission properties. Hot Jupiter
atmospheric modeling efforts would greatly benefit from this
information if it were to become available through applications of
eclipse mapping techniques. In an attempt to evaluate the feasibility
of eclipse mapping for hot Jupiters, we have generated ingress/egress
lightcurves expected for a variety of plausible photospheric emission
models and asked whether these models could be differentiated from one
another, on the basis of repeated eclipse measurements with current or
future infrared facilities.  We found that Spitzer has insufficient
photometric accuracy to make efficient use of these eclipse
diagnostics.  Simple scalings indicate that a system at least ten
times brighter than HD189733b, currently the brightest known eclipsing
hot Jupiter, would be needed for Spitzer to yield meaningful results
in this respect.  On the other hand, our preliminary analysis suggests
that JWST has the potential to reveal relatively subtle features at
the photospheres of currently known eclipsing hot Jupiters.

Our analysis is voluntarily simple and limited in scope. By
construction, we only compare specific photospheric emission
models. This necessarily results in conclusions which are rather
model-dependent. In addition, our treatment of atmospheric radiative
transfer, and the related planet's photospheric emission, was
relatively simple.  We assumed the existence of a well-defined
photospheric surface in all our spectral models, we combined simple
horizontal photospheric temperature maps with the results of 1D
vertical radiative transfer calculations obtained independently and
thus largely ignored important two- or three-dimensional geometrical
effects in our analysis.  Nevertheless, comparisons with crude
bolometric blackbody emission models suggest that our main conclusions
on the feasibility of using eclipse diagnostics with Spitzer and JWST
are not too strongly affected by our various model specifics or
shortcomings, because they mostly depend on the photometric accuracy
achievable for the ingress/egress portions of infrared secondary
eclipses.

Throughout our analysis, we stressed the advantages of non-parametric
inversion methods for eclipse mapping, which would offer
model-independent constraints on the photospheric emission properties
of eclipsing hot Jupiters. By combining data at different wavelengths,
and thus probing photospheres at different heights, 3-dimensional
emission maps of hot Jupiter atmospheres could potentially be
obtained. It remains to be seen with what precision this concept can
be applied in practice.  It is likely that our model-dependent
analysis overestimates the level of details that can be probed by JWST
with non-parametric eclipse mapping methods, but the promising nature
of results obtained so far and the large potential value for the
exoplanet science community provide strong motivations for a more
careful assessment of this exciting new possibility.
}

\acknowledgements

We thank Zoltan Haiman and Joe Patterson for useful discussions and an
anonymous referee for comments that helped improve the manuscript.
This work was supported by NASA contract NNG06GF55G, NASA Astrobiology
Institute contract NNA04CC09A and a Spitzer Theory grant. S. Seager
thanks the Carnegie Institution of Washington for support.

\clearpage

\begin{deluxetable}{cccc}
\tablewidth{0pt}
\tablecaption{System parameters adopted}
\tablehead{
\colhead{}  &  \colhead{HD189733}  &  \colhead{HD209458}  &  \colhead{TrES-1}
}
\startdata
distance (pc)  &  19\tablenotemark{a}  &  47\tablenotemark{a}  &  150\tablenotemark{b} \\
R$_*$ (R$_{\odot}$)  &  0.76\tablenotemark{c}  &  1.146\tablenotemark{d}  &  0.83\tablenotemark{e}  \\
T$_*$ (K)  &  5050\tablenotemark{c}  &  6000\tablenotemark{f}  &  5250\tablenotemark{b} \\
R$_p$ (R$_{Jup}$)  &  1.26\tablenotemark{c}  &  1.35\tablenotemark{d}  &  1.04\tablenotemark{e}  \\
T$_p$ (K)  &  1200  &  1440  &  1160 \\
semi-major axis, $a$ (AU)  &  0.0313\tablenotemark{c}  &  0.046\tablenotemark{f}  &  0.0394\tablenotemark{e} \\
orbital period, $P$ (days)  &  2.219\tablenotemark{c}  &  3.524\tablenotemark{f}  &  3.030\tablenotemark{b}  \\
orbital velocity, $v$ (km s$^{-1}$)  &  154  &  142  &  142  \\
orbital inclination, $i$ (\degrees)  &  85.3\tablenotemark{c}  &  87.1\tablenotemark{g}  &  89.5\tablenotemark{e} \\
star-planet offset, $h$ (R$_{p}$)  &  4.28  &  3.62  &  0.69 \\
ingress/egress $t_{\rm gress}$ (s)  &  1770  &  1520  &  1050 \\
\enddata
\tablerefs{$^\mathrm{a}$\citet{Hipp}, $^\mathrm{b}$\citet{Alonso2004}, $^\mathrm{c}$\citet{Bouchy2005}, $^\mathrm{d}$\citet{Brown2001}, $^\mathrm{e}$\citet{Sozzetti2004}, $^\mathrm{f}$\citet{Mazeh2000}, $^\mathrm{g}$\citet{Charbonneau2000}}
\end{deluxetable}

\begin{deluxetable}{cccc}
\tablewidth{0pt} \tablecaption{JWST and Spitzer single-measurement
errors, $\sigma_1$, and numbers of measurements, $N_1$, during a full
ingress/egress} \tablehead{ \colhead{Instrument,} &
\colhead{HD189733b} & \colhead{HD209458b} & \colhead{TrES-1} \\
\colhead{wavelength (\micron)} & \colhead{$\sigma_1$,\ $N_1$} &
\colhead{$\sigma_1$,\ $N_1$} & \colhead{$\sigma_1$,\ $N_1$} }
\startdata JWST, G140H, 1.4 & 1.5, 536 & \emph{1.4}, 461 & 26, 318
\\ JWST, G235H, 2.3 & 0.3, 536 & \emph{0.4}, 461 & 4, 318 \\ JWST,
G395H, 4 & 0.02, 393 & \emph{0.04}, 338 & 0.3, 233 \\ \hline
Spitzer, IRAC, 4.5 & 0.29, 134 & 0.64, 115 & \textbf{4.1}, 80 \\
Spitzer, IRAC, 8 & 0.27, 134 & 0.71, 115 & \textbf{3.8}, 80 \\ Spitzer,
IRS, 16 & \textbf{0.46}, 120 & 1.3, 103 & 6.2, 71 \\ Spitzer, MIPS, 24
& 1.2, 144 & \textbf{3.5}, 124 & 17, 85 \\ \enddata
\tablecomments{Values in bold are calculated from previous
measurements and values in italics are calculated from the
estimates of \citet{Valenti2005}.  All other values were scaled
accordingly (as described in Section~\ref{sec:spitzer_errors}).}
\end{deluxetable}

\clearpage

\begin{deluxetable}{cccc}
\rotate
\tablewidth{0pt}
\tablecaption{Number of eclipses needed to differentiate between emission
models with Spitzer or JWST.}
\tablehead{
\colhead{System}  &  \colhead{Instrument, wavelength (\micron)}  &
\colhead{Models compared}  &  \colhead{N$_{eclipses}$\tablenotemark{c}}
}
\startdata
\multicolumn{4}{c}{Spitzer} \\
\hline
any system  & any wavelength  &  RadEq, Uniform  &  $>$500 \\
HD189733  &  IRAC, 8  & RadEq, CS05 w/no timing uncert.  &  39 (78) \\
HD189733  &  bolometric w/8\micron~noise stats.\tablenotemark{a}  & RadEq, CS05
w/no timing uncert.  &  9 (18) \\
HD189733  &  IRAC, 8  & RadEq, CS05 w/7s timing uncert.  &  52 (104) \\
HD189733  &  bolometric w/8\micron~noise stats.\tablenotemark{a} & RadEq, CS05 w/7s timing uncert.
&  10 (21) \\
\cutinhead{JWST}
TrES-1  &  NIRSpec, any grating  & RadEq, Uniform  & $>$500 \\
HD209458  &  NIRSpec, 1.4 or 2.3  & RadEq, Uniform  &  $>$500 \\
HD209458  &  NIRSpec, 4  &  RadEq, Uniform  &  2  \\
HD209458  &  bolometric w/4\micron~noise stats.\tablenotemark{a}  &  RadEq, Uniform  &  1 \\
HD209458  &  NIRSpec, 1.4 & RadEq, CS05 w/7s or w/no timing uncert.  &
$>$450 \\
HD209458  &  NIRSpec, 2.3 & RadEq, CS05 w/7s or w/no timing uncert.  &  3 \\
HD209458  &  NIRSpec, 4 & RadEq, CS05 w/7s or w/no timing uncert.  &  1 \\
HD189733  &  NIRSpec, 1.4  &  RadEq, any model  &  $>$500 \\
HD189733  &  NIRSpec, 2.3  &  RadEq, Uniform  &  $>$300 \\
HD189733  &  NIRSpec, 2.3  &  RadEq, CS05 w/7s or w/no timing uncert.  &  2 (5)
\\
HD189733  &  bolometric w/2.3\micron~noise stats\tablenotemark{a}.  &  RadEq, CS05  &  3 (11) \\
HD189733  &  NIRSpec, 4  &  RadEq, any model  &  1  (1)\\
HD189733  &  NIRSpec, 4  &  RadEq, Uniform w/ld\tablenotemark{b}  &  1 (1)\\
HD189733  &  NIRSpec, 4  &  Uniform, Cho03  &  2  (2)\\
HD189733  &  bolometric w/4\micron~noise stats.\tablenotemark{a}  &  RadEq, any model  &  1 (1)\\
HD189733  &  bolometric w/4\micron~noise stats.\tablenotemark{a}  &  Uniform, Cho03  &  1  (1)\\
\enddata

\tablenotetext{a}{In the highly idealized ``bolometric'' models,
ingress/egress curves are calculated assuming spectrally-integrated
blackbody emission from the planet's various photospheric surface
elements.  Cadences and noise properties used to simulate eclipse data
from these bolometric models are the same as the instrument-specific
values assumed in all other models.}  \tablenotetext{b}{Model with
additional limb darkening imposed.}\tablenotetext{c}{{The
number of eclipses in parenthesis is for the value of HD189733b's
radius recently revised down by 10\%. See Appendix~A for details.}}

\end{deluxetable}
\clearpage

\appendix

\section{Revised HD189733b Radius}

{Recently, \citet{Bakos06} and \citet{Winn07} have revised the
value of the radius of HD189733b downward from $1.26$ to $1.15$
Jupiter radii. Accordingly, we have reconsidered our eclipse
lightcurve analysis for this planet. The revised number of eclipses
needed to differentiate between various photospheric emission models
appears in parenthesis in Table~3. The smaller radius results in
increased photometric errors and thus an increased number of required
eclipses. Nonetheless, we find that the photometric accuracy of JWST
at 4\micron \ remains good enough for one or two eclipses to
distinguish reliably among the various emission models considered.}

\clearpage

\begin{thebibliography}

\bibitem[Agol et al.(2005)]{Agol2005} Agol, E., Steffen, J., Sari, R., \& Clarkson, W.\ 2005, \mnras, 359, 567 

\bibitem[Alonso et al.(2004)]{Alonso2004} Alonso, R., et al.\ 2004, \apjl, 613, L153 

\bibitem[Baptista(2001)]{baptista01} Baptista, R. 2001, in
Astrotomography, Indirect Imaging Methods in Observational Astronomy,
Eds H.M.J. Boffin, D. Steeghs and J. Cuypers, Lecture Notes in
Physics, vol. 573, p.307

\bibitem[Bakos et al.(2006)]{Bakos06} Bakos, G.~A. et al. \ 2006, ApJ,
650, 1160

\bibitem[Barman et al.(2005)]{Barman2005} Barman, T.~S., Hauschildt, P.~H., \& Allard, F.\ 2005, \apj, 632, 1132 

\bibitem[Bobinger(2000)]{bobinger00} Bobinger, A. 2000, A\&A, 357, 1170

\bibitem[Bouchy et al.(2005)]{Bouchy2005} Bouchy, F., et al.\ 2005, \aap, 444, L15

\bibitem[Brown et al.(2001)]{Brown2001} Brown, T.~M., Charbonneau, D., Gilliland, R.~L., Noyes, R.~W., \& Burrows, A.\ 2001, \apj, 552, 699

\bibitem[Burkert et al.(2005)]{Burkert2005} Burkert, A., Lin, D.~N.~C., Bodenheimer, P.~H., Jones, C.~A., \& Yorke, H.~W.\ 2005, \apj, 618, 512

\bibitem[Burrows et al.(2005)]{Burrows2005} Burrows, A., Hubeny, I., \& Sudarsky, D.\ 2005, \apjl, 625, L135 

\bibitem[Burrows et al.(2006)]{Burrows2006} Burrows, A., Sudarsky, D., \& Hubeny, I.\ 2006, ApJ, 650, 1140

\bibitem[Butler et al.(2006)]{Butler06} Butler, R.~P. et al. 2006,
ApJ, 646, 505

\bibitem[Charbonneau et al.(2005)]{Charbonneau2005} Charbonneau, D., et al.\ 2005, \apj, 626, 523 

\bibitem[Charbonneau et al.(2000)]{Charbonneau2000} Charbonneau, D., Brown, T.~M., Latham, D.~W., \& Mayor, M.\ 2000, \apjl, 529, L45

\bibitem[Cho et al.(2003)]{Cho} Cho, J.~Y-K., Menou, K., Hansen, B.~M.~S., \& Seager, S.\ 2003, \apjl, 587, L117 

\bibitem[Cho et al.(2006)]{Cho06} Cho, J.~Y-K., Menou, K., Hansen, B.,
\& Seager, S.\ 2006, ApJ submitted, astro-ph/0607338

\bibitem[Cooper \& Showman(2005)]{C&S} Cooper, C.~S., \& Showman, A.~P.\ 2005, \apjl, 629, L45

\bibitem[de Pater \& Lissauer(2001)]{dePater2001} de Pater, I., \& Lissauer, J.\ 2001, Planetary Sciences, by Imke de Pater and Jack Lissauer.~Cambridge University Press, 2001, 544 pp. 

\bibitem[Deming et al.(2006)]{Deming2006} Deming, D., Harrington, J., Seager, S., \& Richardson, L.~J.\ 2006, \apj, 644, 560

\bibitem[Deming et al.(2005)]{Deming2005} Deming, D., Seager, S., Richardson, L.~J., \& Harrington, J.\ 2005, \nat, 434, 740

\bibitem[Harrington et al.(2006)]{HarrUps06} Harrington, J. et
al. 2006, Science, 314, 623

\bibitem[Holman \& Murray(2005)]{HolMur05} Holman, M.~J. \& Murray,
N.~W. 2005, Science, 307, 1288

\bibitem[Horne(1985)]{horne85} Horne, K. 1985, MNRAS, 213, 129

\bibitem[Iro et al.(2005)]{Iro} Iro, N., Bezard, B. \& Guillot, T
2005, A\&A, 436, 719

\bibitem[Fortney et al.(2005)]{Fortney2005} Fortney, J.~J., Marley, M.~S., Lodders, K., Saumon, D., \& Freedman, R.\ 2005, \apjl, 627, L69 

\bibitem[Fortney et al.(2006)]{Fortney2006} Fortney, J.~J., Cooper,
C.~S., Showman, A.~P., Marley, M.~S.\& Freedman, R.\ 2006, \apj~in
press, astro-ph/0608235

\bibitem[Lubow et al.(1997)]{Lubow1997} Lubow, S.~H., Tout, C.~A., \& Livio, M.\ 1997, \apj, 484, 866

\bibitem[Mazeh et al.(2000)]{Mazeh2000} Mazeh, T., et al.\ 2000, \apjl, 532, L55
\bibitem[Menou et al.(2003)]{Menou03} Menou, K. Cho, J.~Y.-K., Seager,
S. \& Hansen, B.~M.~S.\ 2003, \apjl, 587, L113

\bibitem[Ogilvie \& Lin(2004)]{Ogilvie2004} Ogilvie, G.~I., \& Lin, D.~N.~C.\ 2004, \apj, 610, 477 

\bibitem[Perryman et al.(1997)]{Hipp} Perryman, M.~A.~C., et al.\ 1997, \aap, 323, L49

\bibitem[Rasio et al.(1996)]{Rasio1996} Rasio, F.~A., Tout, C.~A., Lubow, S.~H., \& Livio, M.\ 1996, \apj, 470, 1187

\bibitem[Rutten(1996)]{rutten96} Rutten, R.~C.~M. 1996, in proceedings
of the 176th Symposium no. 176 of the International Astronomical
Union, Eds Klaus G. Strassmeier and Jeffrey L. Linsky, Kluwer Academic
Publishers, Dordrecht, p.69

\bibitem[Rutten(1998)]{rutten98} Rutten, R.~C.~M. 1998, A\&AS, 127, 581

\bibitem[Seager et al.(2005)]{Seager2005} Seager, S., Richardson, L.~J., Hansen, B.~M.~S., Menou, K., Cho, J.~Y.-K., \& Deming, D.\ 2005, \apj, 632, 1122

\bibitem[Showman \& Guillot(2002)]{ShowGui} Showman, A.~P. \& Guillot,
T. 2002, A\&A, 385, 166

\bibitem[Sozzetti et al.(2004)]{Sozzetti2004} Sozzetti, A., et al.\ 2004, \apjl, 616, L167

\bibitem[Spruit(1994)]{spruit94} Spruit, H.~C. 1994, A\&A, 289, 441

\bibitem[Stern(1992)]{Stern1992} Stern, S.~A.\ 1992, \araa, 30, 185

\bibitem[Valenti et al.(2005)]{Valenti2005} Valenti, J.~A., et al.\
2005, BAAS, 37, 1350

\bibitem[Warner et al.(1971)]{Warner1971} Warner, B., Robinson, E.~L., \& Nather, R.~E.\ 1971, \mnras, 154, 455

\bibitem[Williams et al.(2006)]{Williams2006} Williams, P.~K.~G., Charbonneau, D., Cooper, C.~S., Showman, A.~P., \& Fortney, J.~J.\ 2006, \apj, 649, 1020

\bibitem[Winn et al.(2007)]{Winn07} Winn, J.~N. et al. \ 2007, AJ,
133, 1828

\end{thebibliography}
\end{document}